\documentclass[12pt]{article}
\usepackage{a4wide}
\usepackage{graphicx}
\usepackage[centertags]{amsmath}
\usepackage{amssymb}
\usepackage{cite}
\usepackage{enumerate}
\usepackage{scrtime}
\usepackage{color}
\usepackage[sans]{dsfont} 
\usepackage{colortbl}
\usepackage{subfigure}
\usepackage{paralist}
\usepackage{axodraw}
\usepackage{lscape}
\usepackage{longtable}
\usepackage{ulem}
\usepackage{cancel}

\graphicspath{{./}{./graphs/}}

\newcommand{\U}[1]{\ensuremath{\mathrm{U}(#1)}}

\newcommand{\mex}{{M_{\textsc{ex}}}}

\newcommand{\mstring}{M_{\textsc{s}}}

\def\z6ii{$\mathbb{Z}_6$-II}

\def\beqn{\begin{eqnarray}}
\def\eeqn{\end{eqnarray}}
\def\bsqn{\begin{subequations}}
\def\esqn{\end{subequations}}

\def\bctr{\begin{center}}
\def\ectr{\end{center}}

\begin{document}

\thispagestyle{empty}

\begin{flushright}
OHSTPY-HEP-T-08-005  \\
\end{flushright}
\vskip 1cm

\begin{center}
{\textit{Addendum to Reconciling Grand Unification with Strings\\
by Anisotropic Compactifications}} \vskip1cm
\end{center}
\noindent \centerline{\bf Ben Dundee, Stuart Raby} \vskip1cm
\centerline{ \em Department of Physics, The Ohio State University,}
\centerline{\em 191 W.~Woodruff Ave, Columbus, OH 43210, USA}
\vskip1cm \centerline{\bf Ak\i{}n Wingerter} \vskip1cm
\centerline{\em Indian Institute of Science,} \centerline{\em
Bangalore 560 012, INDIA } \vskip1cm

In a recent paper \cite{Dundee:2008ts}, by working in the orbifold
GUT limit of the Heterotic string, we showed how one could
accommodate gauge coupling unification in the ``mini-landscape''
models of References \cite{Buchmuller:2005jr, Buchmuller:2006ik,
Lebedev:2006tr, Lebedev:2006kn, Lebedev:2007hv}.  Furthermore, it
was shown how one of the solutions was consistent with the
decoupling of other exotics and $F=0$.  In this short addendum, we
show that this solution is also consistent with $D=0$.

Let us first describe the steps one must take to show that there is
a solution to the equations  $F = D = 0$.   For simplicity, and
without loss of generality, we will consider a $\U1_A \times \U1_B$
gauge theory. We will further consider $N$ fields $\Phi_i$ charged
under both $\U1$s, where each $\Phi_i$ has charge $q_i^A$ under the
first \U1, and charge $q_i^B$ under the second \U1.  If we turn the
superpotential off, unbroken supersymmetry (SUSY) requires
\begin{eqnarray} \label{first_d_eq_zero}
 D_A &\equiv& \sum_{i=1}^N q_i^A \left|\Phi_i\right|^2 = 0 \\ \label{second_d_eq_zero}
 D_B &\equiv& \sum_{i=1}^N q_i^B \left|\Phi_i\right|^2 = 0.
\end{eqnarray}

It is well known that the moduli space of $D=0$ is spanned by a basis of holomorphic,
gauge invariant monomials (HIMs) \cite{Cleaver:1997jb}.  Quite generally, the dimension of the
moduli space $\mathcal{M}$ of some gauge group $\mathcal{G}$ is given by the number
of fields charged under $\mathcal{G}$ minus the number of constraints coming from
$V_D = 0$:
\begin{equation}
 {\rm dim~}\mathcal{M} = N - {\rm dim~}\mathcal{G}.
\end{equation}

The HIMs can be represented as vectors $\vec{x}^\alpha$ in the
$\U1_A \times \U1_B$ charge space.  That is, if we define the charge
matrix $\mathbb{Q}$ as
\begin{equation}
 \mathbb{Q} \equiv \left( \begin{array}{cccc} q_1^A&q_2^A&\cdots&q_N^A\\
q_1^B&q_2^B&\cdots&q_N^B \end{array} \right),
\end{equation}
then the set of HIMs are defined by the solutions to
\begin{equation} \label{null_space}
 \mathbb{Q}\cdot \vec{x}^\alpha = 0.
\end{equation}
The requirement that the monomials be holomorphic is a non-trivial
constraint---effectively, this means that the entries in
$\vec{x}^\alpha$ be positive semi-definite integers.\footnote{Here
we will note that it is entirely possible that the null space of the
charge matrix $Q$ is empty---that is, it could very well be that
there exists no holomorphic, gauge invariant monomials. This
corresponds to a situation where SUSY is broken spontaneously
\textit{everywhere} in moduli space by $D$ terms, except possibly at
the origin where one would expect an enhanced gauge symmetry.}  The
set of $\vec{x}^\alpha$s are linearly independent N dimensional
vectors spanning a space with $ {\rm dim~}\mathcal{M} = N - {\rm
dim~}\mathcal{G}$.   Then the set of monomials spanning the moduli
space $\cal{H}$ is given by
\begin{equation}
 \mathcal{H} = \{ M_{\alpha} =
 \Phi_1^{x_1^{\alpha}}\Phi_2^{x_2^{\alpha}}\cdots\Phi_N^{x_N^{\alpha}} \}.
\end{equation}

We can guarantee solutions to $V_D = 0$ if we demand that
\begin{equation} \label{condition_for_phis}
 \left|\Phi_i\right|^2 = \Phi_i \frac{\partial}{\partial \Phi_i}\sum_{\alpha}^{{\rm dim~}\mathcal{M}} a_{\alpha} M_{\alpha}.
\end{equation}
In general, one must choose the constants $a_{\alpha}$ such that all of the phases
on the right hand side of Equation (\ref{condition_for_phis}) cancel.  A substitution into
Equations (\ref{first_d_eq_zero}) and (\ref{second_d_eq_zero}) shows that we do indeed satisfy
$D = 0$.

Next, consider the case where we turn on a superpotential,
$\mathcal{W}$.  The superpotential is an arbitrary function of the
holomorphic, gauge invariant monomials given by $\mathcal{W}=
\mathcal{W}(M_{\alpha})$.  The requirement that $F=0$ can be stated
as follows:
\begin{equation}
 \Phi_i \frac{\partial}{\partial \Phi_i}\mathcal{W}(M_{\alpha}) = 0.
\end{equation}
This only tells us something that we already knew---the superpotential only constrains
combinations of the holomorphic, gauge invariant monomials $M_{\alpha}$, not the fields
themselves.\footnote{See Chapter VIII in Reference \cite{Wess:1992cp}, for example.}  We can
solve these constraints explicitly for the $M_{\alpha}$, and then express $|\Phi_i|^2$ in
terms of a linear combination of the $M_{\alpha}$ (as before), with arbitrary $a_{\alpha}$.
Thus we see that it is \textit{always} possible to satisfy $D=0$ when given a solution to
$F=0$.\footnote{A different argument was made by Luty and Taylor \cite{Luty:1995sd}.}

Finally we consider the (relevant) case where the ``$A$'' in $\U1_A \times \U1_B$ stands
for ``anomalous''.  In this case, Equation (\ref{first_d_eq_zero}) is modified slightly.
We must now cancel a Fayet-Iliopolous (FI) term if we wish to keep SUSY unbroken:
\begin{equation}
 D_A = \sum_i \left| \Phi_i \right|^2 + \left|\xi\right| = 0.
\end{equation}
In order to ensure that there exists a direction in moduli space along which this constraint
can be satisfied, we seek at least one HIM which has a net negative charge under $\U1_A$.  This
will ensure that we can cancel the (negative) FI term.

We now turn to the issue which we would like to address, namely
proving $D=0$ for the solution presented in Section 4 of Reference
\cite{Dundee:2008ts}, wherein it was shown that $F=0$ for one of the
models, but the issue of $D=0$ was neglected.  We consider Model 1A
in Reference \cite{Lebedev:2007hv}, where it was shown that
solutions to $F=D=0$ existed for arbitrary (string scale) vevs for
some subset of the non-Abelian singlet fields---see Equation (5.3)
of \cite{Lebedev:2007hv}.  However for gauge coupling unification we
require two fields, called $s_1$ and $s_{25}$, to have intermediate
scale ($\mex$) vevs, while several other fields are required to have
vevs of order the string scale \cite{Dundee:2008ts}. There are also
several other non-Abelian singlet fields which we require to get
zero vevs.  (The complete list of fields, along with their charges
are listed in Appendix E of Reference \cite{Lebedev:2007hv}.)

Using the arguments above, we note that the proof of $D=0$ for our
solution is straightforward.  One only need check that there are
enough HIMs, including all fields, to saturate the dimension of the
moduli space, and that there exists at least one holomorphic
monomial, excluding the fields $s_1$ and $s_{25}$, which has a
negative charge under the $\U1_A$. We have verified that this is the
case.

In closing, we will note that simply taking $s_1$ and $s_{25}$ out
of the charge matrix $\mathbb{Q}$ produces a null result for
$\mathbb{Q}\cdot\vec{x}^\alpha = 0$, meaning that there are no
vectors $\vec{x}^\alpha$ which satisfy the above equation if we set
the vevs of $s_1 = s_{25} = 0$.  We point out that the solution to
$F=0$ does require one engineer a cancellation on the order $\mex /
\mstring \sim 10^{-8}$ among the other vevs.  We should expect,
then, that a tuning in the coefficients $a_{\alpha}$ is required to
this order as well.  While this is aesthetically unappealing, it is
nonetheless possible, as the relationship in Equation
(\ref{condition_for_phis}) only constrains the phases of the
$a_{\alpha}$s and not their magnitudes.

\noindent {\bf Acknowledgement}

This work is partially supported by DOE grant DOE/ER/01545-881. B.D.
and S.R. also thank the Stanford Institute for Theoretical Physics
for their hospitality.

\clearpage
\bibliographystyle{utphys}
\bibliography{mybibliography}

\providecommand{\href}[2]{#2}\begingroup\raggedright\begin{thebibliography}{10}

\bibitem{Dundee:2008ts}
  B.~Dundee, S.~Raby and A.~Wingerter,
  ``Reconciling Grand Unification with Strings by Anisotropic
  Compactifications,''
  arXiv:0805.4186 [hep-th].

\bibitem{Buchmuller:2005jr}
  W.~Buchm{\"{u}}ller, K.~Hamaguchi, O.~Lebedev and M.~Ratz,
  ``Supersymmetric standard model from the heterotic string,''
  Phys.\ Rev.\ Lett.\  {\bf 96}, 121602 (2006)
  [arXiv:hep-ph/0511035].
  
\bibitem{Buchmuller:2006ik}
  W.~Buchm{\"{u}}ller, K.~Hamaguchi, O.~Lebedev and M.~Ratz,
  ``Supersymmetric standard model from the heterotic string. II,''
  Nucl.\ Phys.\  B {\bf 785}, 149 (2007)
  [arXiv:hep-th/0606187].

\bibitem{Lebedev:2006tr}
O.~Lebedev {\em et al.}, ``Low energy supersymmetry from the heterotic
  landscape,'' {\em Phys. Rev. Lett.} {\bf 98} (2007) 181602,
\href{http://www.arXiv.org/abs/hep-th/0611203}{{\tt hep-th/0611203}}.

\bibitem{Lebedev:2006kn}
O.~Lebedev {\em et al.}, ``A mini-landscape of exact MSSM spectra in heterotic
  orbifolds,'' {\em Phys. Lett.} {\bf B645} (2007) 88--94,
\href{http://www.arXiv.org/abs/hep-th/0611095}{{\tt hep-th/0611095}}.

\bibitem{Lebedev:2007hv}
O.~Lebedev {\em et al.}, ``The heterotic road to the mssm with $R$ parity,''
\href{http://www.arXiv.org/abs/arXiv:0708.2691 [hep-th]}{{\tt arXiv:0708.2691
  [hep-th]}}.

\bibitem{Cleaver:1997jb}
  G.~Cleaver, M.~Cvetic, J.~R.~Espinosa, L.~L.~Everett and P.~Langacker,
  ``Classification of flat directions in perturbative heterotic superstring
  vacua with anomalous U(1),''
  Nucl.\ Phys.\  B {\bf 525}, 3 (1998)
  [arXiv:hep-th/9711178].

\bibitem{Wess:1992cp}
  J.~Wess and J.~Bagger,
  \textit{Supersymmetry and supergravity},
{\it  Princeton, USA: Univ. Pr. } (1992)

\bibitem{Luty:1995sd}
  M.~A.~Luty and W.~Taylor,
  ``Varieties of vacua in classical supersymmetric gauge theories,''
  Phys.\ Rev.\  D {\bf 53}, 3399 (1996)
  [arXiv:hep-th/9506098].




  
\end{thebibliography}\endgroup

\end{document}